\documentclass[aps,prl,reprint,superscriptaddress,floatfix,amsmath,showpacs]{revtex4-1}

\usepackage{graphicx}% Include figure files
\usepackage{amsmath}
\usepackage{textcomp} %texterweiterungen

\begin{document}

\title{Storage-recovery phenomenon in magnonic crystal}

\author{A. V. Chumak}
\email{chumak@physik.uni-kl.de}
\affiliation{Fachbereich Physik and Forschungszentrum OPTIMAS, Technische Universit\"at
Kaiserslautern, 67663 Kaiserslautern, Germany}

\author{V. I. Vasyuchka}
\affiliation{Fachbereich Physik and Forschungszentrum OPTIMAS, Technische Universit\"at Kaiserslautern, 67663
Kaiserslautern, Germany}

\author{A. A. Serga}
\affiliation{Fachbereich Physik and Forschungszentrum OPTIMAS, Technische Universit\"at Kaiserslautern, 67663
Kaiserslautern, Germany}

\author{M. P. Kostylev}
\affiliation{School of Physics, University of Western Australia, Crawley, Western Australia 6009, Australia}

\author{B. Hillebrands}
\affiliation{Fachbereich Physik and Forschungszentrum OPTIMAS, Technische Universit\"at Kaiserslautern, 67663 Kaiserslautern, Germany}

\date{\today}

\begin{abstract}

The phenomenon of wave trapping  in an artificial crystal with limited number of periods is demonstrated experimentally
using spin waves in a magnonic crystal. The information stored in the crystal is recovered afterwards by parametric
amplification of the trapped wave. The storage process is based on the excitation of standing internal crystal modes and
differs principally from the well-known phenomenon of deceleration of light in photonic crystals.

\end{abstract}

\maketitle

The deceleration or even the full stop of light due to the modification of the light dispersion in photonic crystals
(PCs) has been a topic of intense experimental and theoretical studies over the last decade \cite{PC1, PC2, PC3, PC4}.
A wave of light propagating through a PC couples with the internal standing PC mode and generates a slow light mode. In
terms of the dispersion characteristics this means that the slope of the dispersion curve decreases at the edges of the
band gaps which results in an extremely small value of the group velocity. It has been demonstrated that the slow light
can be used for time-domain processing including buffering (storage and recovery) of optical signals as well as for an
enhancement of nonlinear effects due to the spatial compression of optical energy \cite{PC3, PC4}.

Magnonic crystals (MCs) are the magnetic counterpart of photonic crystals which operate with spin waves, i.e. the
collective oscillations of the spin lattice of a magnetic material \cite{MC-Nikitov, MC-Kruglyak, MC-Chumak-BVMSW,
MC-Ustinov, MC-Puszkarski, MC-Lee, MC-Chumak-muMC, MC-Wang, MC-review-Kostylev}. A wide range of parameters, which
determine the spin-wave characteristics, can be periodically modulated to form the MC. This, as well as the possibility
of fast dynamic control of these parameters \cite{MC-Chumak-DMC1, MC-Chumak-DMC2} make MCs promising candidates for the
transfer and processing of information in the GHz frequency range. However, in spite of the considerable recent
progress in MCs studies, neither spin-wave deceleration nor storage-recovery of a spin-wave carried signal has been
demonstrated yet. This may be due to the spin-wave damping which allows fabrication of MCs comprising usually not more
than 20 periods (see Ref.~\cite{YIG-magnonics} and references therein). The small number of periods implies that the
slope of MC dispersion does not become zero at the edges of the magnonic band gap in contrast to light dispersion at
the edge of photonic gaps. Hence, rather than vanishing, the group velocity of spin waves only slightly decreases at
the gap edges. This makes the realization of storing a signal in a MC using the \emph{slow} spin-wave mode
questionable.

\begin{figure}[h]
\includegraphics[width=8 cm]{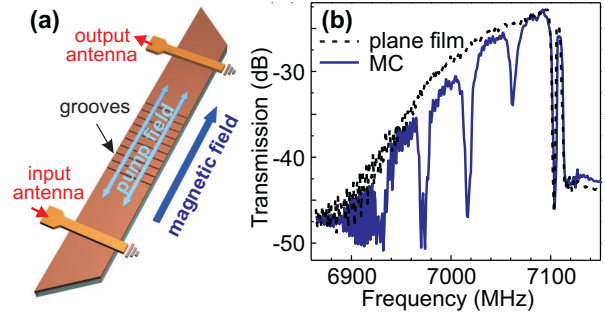}
\caption{\label{Fig1} (Color online) (a) Sketch of the experimental setup. The MC is fabricated in the form of YIG
spin-wave waveguide with 10 grooves on its surface. The Rf pump magnetic field is parallel to the bias field $H$. (b)
Measured transmission characteristics (the ratio of the output signal intensity to the input) of the plane YIG film and
MC for the magnetic field $H = 1800$~Oe. Several band gaps are seen.}
\end{figure}

As an alternative solution, in this Letter we show that the storage-recovery phenomenon can be successfully realized in
an artificial crystal with a limited number of periods by the use of an internal \emph{standing} mode of the crystal.
This mode is excited by the incident spin wave and conserves oscillation energy after the propagating wave has left the
MC area. Upon subsequent parametric amplification the internal mode irradiates part of its energy back into the
propagating spin wave allowing the coherent restoration of the stored microwave signal. The restoration occurs at the
edges of the magnonic crystal band gaps in narrow (1.2~MHz) frequency windows coinciding with the local minima of the
spin-wave group velocity.

The magnonic crystal used in our experiment had been produced in the form of a stripe of a low-damping magnetic
insulator (5.1~$\mu$m-thick yttrium iron garnet (YIG) film) with an array of parallel grooves chemically etched into
its surface (see Fig.~\ref{Fig1}(a)) \cite{MC-Chumak-BVMSW}. The array comprises ten 300~nm-deep and 30~$\mu$m-wide
grooves placed 270~$\mu$m apart (lattice constant $a=300~\mu$m). The bias magnetic field is applied along the stripe in
order to arrange conditions for propagation of backward volume magnetostatic spin waves (BVMSW) \cite{YIG-magnonics}.
These waves were previously found to be excellent signal carriers for one-dimensional magnonic crystals. For example,
it has been shown that even small regular distortions of the surface of a magnetic film result in the appearance of
pronounced rejection bands in the BVMSW frequency spectrum  \cite{MC-Chumak-JAP}. The waves were excited and detected
in the YIG film waveguide using microwave stripline antennas placed at equal distances from both ends of the grooved
area and 8~mm apart from each other (Fig.~\ref{Fig1}(a)). Microwave power of 0.3~mW applied to the input antenna was
sufficiently low to avoid any non-linear effects which could potentially influence the input spin wave and storage
process. The measured transmission characteristics for the MC along with the one for the reference unstructured YIG
waveguide are shown in the Fig.~\ref{Fig1}(b). Several band gaps where spin waves are not able to propagate are clearly
visible.

The amplification of the signals stored in the magnonic crystal was realized by means of pulsed parallel
electromagnetic pumping. Quantum-mechanically such a pumping represents a three-particle process \cite{Gurevich, Lvov},
in which one photon of the pumping electromagnetic field splits into two magnons having half the pumping frequency and
opposite wavevectors. This phenomenon has been successfully used for spin-wave amplification, wave-front reversal, and
for the recovery of a microwave signal stored as a thickness spin-wave mode of a ferromagnetic film (see review
\cite{YIG-magnonics}). In our experiment a pumping microwave magnetic field oriented parallel to the bias magnetic
field (see Fig.~\ref{Fig1}(a)) was concentrated in the grooved area of the YIG stripe by a dielectric resonator
\cite{Neumann-DR} having a resonance frequency of $14.424$~GHz.

\begin{figure}[h]
\includegraphics[width=8 cm]{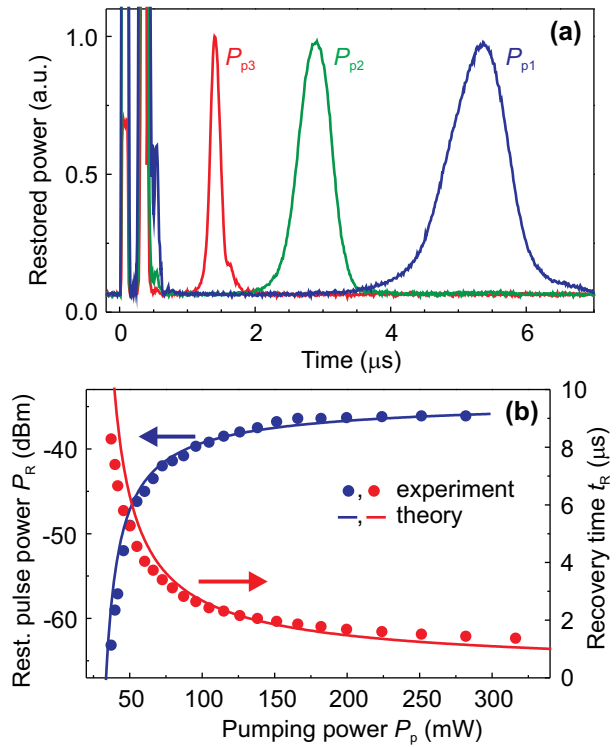}
\caption{\label{Fig2} (Color online) (a) The normalized time profiles of the restored signal measured at field $H = 1860$~Oe for different pumping powers: $P_\mathrm{p1} = 50$~mW, $P_\mathrm{p2} = 90$~mW, $P_\mathrm{p3} = 320$~mW. (b) Measured and calculated dependencies of the restored signal power and recovery time as a function of the pumping power.}
\end{figure}

The storage-recovery experiment has been performed in the following way. A 100~ns-long microwave pulse of $f_\mathrm{s}
= 7.212$~GHz frequency is applied to the input antenna in order to excite a traveling spin-wave packet which propagates
toward the output antenna. The time traces of the output signal are shown in Fig.~\ref{Fig2}(a). First, the output
antenna receives a practically rectangular pulse without any delay, caused by a direct electromagnetic leakage from the
input antenna at the time $t = 0$. Approximately $0.3~\mu$s afterwards the pulsed signal was delivered to the output
antenna by the traveling packet of the initially excited spin waves. Well after this took place we applied a
10~$\mu$s-long pumping pulse at the frequency of $f_\mathrm{p} = 2f_\mathrm{s} = 14.424$~GHz. This resulted in the
appearance of an additional bell-shaped pulse at the output antenna. This is the restored signal.

The restoration mechanism used in our experiments is very common and has been already used to recover the signals
stored in standing thickness spin-wave modes in plane magnetic films \cite{Momentum_relaxation-PRL}. It is based on
frequency selective parametric amplification of a stored wave or oscillation, which can have any physical nature. In a
frequency degenerated multi-mode system this artificially excited stored mode is amplified simultaneously with modes
from the thermal bath. The competitive interaction between these modes and the consequent process of saturation of the
parametric amplification are responsible for the restored pulse characteristics. A qualitative analytical theory of the
restoration process of this type is given in \cite{Restor-PRL}, while a quantitative theory is presented in
\cite{Restor-PRB, Variable_damping-PRB}.

The measured  and calculated dependencies of the restored signal power $P_\mathrm{R}$ and of the recovery time
$t_\mathrm{R}$ on the pumping power $P_\mathrm{p}$ are shown in Fig.~\ref{Fig2}(b). The presented data have been
obtained with a delay between the applications of the input signal and the pump pulse of $0.5~\mu$s (no qualitative
difference was observed when this time delay was varied between $0.4~\mu$s and $1.4~\mu$s). The calculation has been
performed using the analytical formulas from Ref.~\cite{Restor-PRL} for the same parameters (the relaxation frequency
of the thermal mode is 3.7~MHz, of the stored mode 4.3~MHz, the difference between the thermal amplitude level
$A_\mathrm{T}$ and the critical level of amplification saturation $A_\mathrm{cr}$ is $\ln(A_\mathrm{cr}/A_\mathrm{T}) =
10.25$). It can be seen that the measured and calculated recovery time $t_\mathrm{R}$ decreases and the restored signal
power $P_\mathrm{R}$ increases with an increase in $P_\mathrm{p}$. The reason for this behavior is that an increase in
pumping power results in a stronger parametric amplification but also results in a faster saturation. The excellent
agreement between the theoretical prediction and the experiment gives a solid support to our understanding of the
nature of the restoration process in a magnonic crystal.

\begin{figure}[h]
\includegraphics[width=7.5 cm]{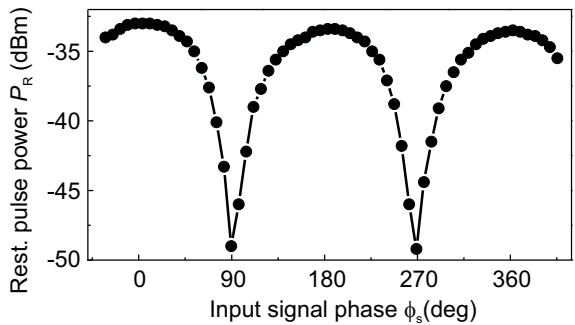}
\caption{\label{Fig3} (Color online) Dependence of the restored signal power on the input signal phase. The signal and
pump generators were locked-in.}
\end{figure}

Completely new and specific for the magnonic crystal is that fact that the standing MC mode, which is used in the
experiment for the signal storage, consists of two counter propagating waves with strictly coupled phases. The phase of
the wave propagating towards the output antenna $\varphi_k$ is determined by the phase of an applied microwave signal
$\varphi_\mathrm{s}$. The phase of the counter-propagating wave $\varphi_{-k}$ is also $\varphi_\mathrm{s}$ and is due
to the reflection from the Bragg lattice. Importantly, the parallel pumping, as a parametric process, also couples two
counter-propagating waves. The sum of the phases of these two modes is defined by the phase of the pump
$\varphi_\mathrm{p}$: $\varphi_k + \varphi_{-k} = \varphi_\mathrm{p} + \pi/2$ \cite{Gurevich, Lvov}. Thus, the
restoration process in a magnonic crystal combines two different mechanisms coupling two counter-propagating waves and
the phase conditions for both these mechanisms should be met simultaneously. As a result the characteristics of the
restored pulse must be influenced by the phase shift between the signal wave and the pumping.

Indeed, we have experimentally registered the strong change of the restored power (more than 15~dB) as a function of
the input signal phase (see Fig.~\ref{Fig3}). The maxima and minima of the restored signal power $P_\mathrm{R}$
correspond to the input signal phase $\varphi_\mathrm{s} = \varphi_\mathrm{p}/2 + \pi/4 + 2\pi n$ and
$\varphi_\mathrm{s} = \varphi_\mathrm{p}/2 - \pi/4+ 2\pi n$, respectively (where $n$ is an integer value). As a result
the phase distance between a neighbour minimum and maximum is exactly $90^\circ$. The observed phase dependence clearly
demonstrates that the signal stored inside the MC is phase correlated to the input microwave signal: in spite of the
distortion of the time profile of the original pulse its phase information is conserved. Furthermore, these results
strongly prove the fact that the internal standing mode of a magnonic crystal participates in the storage-recovery
process.

\begin{figure}[h]
\includegraphics[width=8 cm]{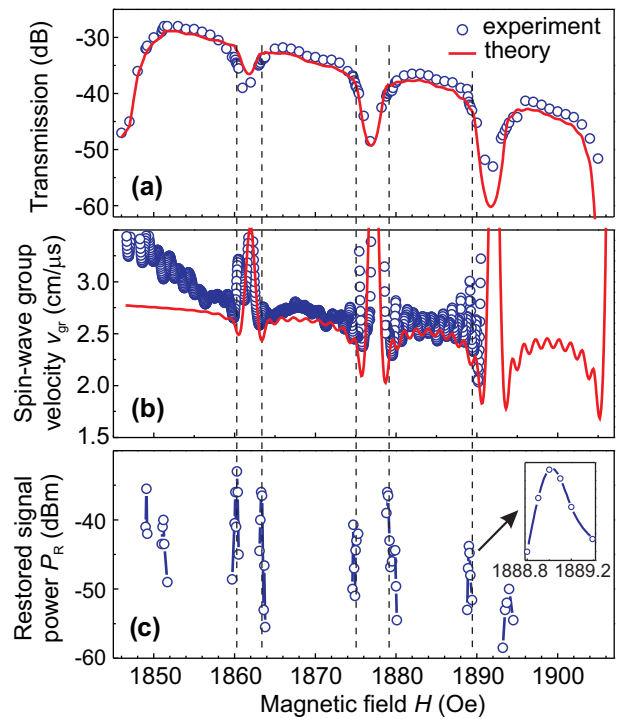}
\caption{\label{Fig4} (Color online) (a)  Transmission of the spin-wave signal as a function of bias magnetic field
(open circles - experiment, line - theory). (b) Dependence of the group velocity $v_\mathrm{gr}$ on the field $H$. The
slight decrease in $v_\mathrm{gr}$ at the edges of the band gap is visible.  (c) Measured restored signal power
$P_\mathrm{R}$ as a function of field $H$. One sees that the restored signal is visible only at the edges of the band
gaps.}
\end{figure}

The power of the  transmitted spin-wave signal, the spin-wave group velocity $v_\mathrm{gr}$, and the power of the
restored signal $P_\mathrm{R}$ are shown in Fig.~\ref{Fig4} as a function of the applied magnetic field. The experiment
has been performed for the fixed signal frequency $f_\mathrm{s} = 7.212$~GHz equal to half of the dielectric resonator
frequency. The variation in the bias field in this case results in the variation of the spin-wave wavenumber, and
therefore Fig.~\ref{Fig4}(a) is practically an reminiscent copy to Fig.~\ref{Fig1}(b) where the transmission
characteristics of the magnonic crystal is shown for a fixed field. We used the transfer matrix approach
\cite{MC-Chumak-BVMSW, MC-Chumak-JAP} to model the field dependence of the spin-wave transmission through the magnonic
crystal. As seen in Fig.~\ref{Fig4}(a), the simulated characteristics are in good agreement with the experiment
(parameters for the simulations were taken from \cite{MC-Chumak-BVMSW}).

We also calculated the spin-wave group velocity $v_\mathrm{gr}$ and compared it with the experimental data
\cite{detail1}. One clearly sees from Fig.~\ref{Fig4}(a) that both the measured and the calculated group velocities
decrease at the edges of the band gaps where the slope of spin-wave dispersion decreases. This decrease is only up to
15 percent and cannot be used in a slowing down approach to store information for a reasonably long time as in the case
of photonic crystals \cite{PC4}.

However, as one can see from Fig.~\ref{Fig4}(c), the maxima of the restored pulse well correlate with the minima of the
group velocity. This means that the recovery process and consequently the signal storage are efficient only at the
edges of the band gaps \cite{detail2}. It is especially surprising that the field width of the regions where the
restoration takes place (see inset in Fig.~\ref{Fig4}(c)) is very narrow (approximately 0.4~Oe) and is comparable with
the ferromagnetic resonance linewidth (0.5~Oe in our case). (Please note that the frequency resolution of the
experiment is given by the spectral width of the pump signal of 0.1~MHz which corresponds to 0.03~Oe on the bias field
scale.)

The appearance of the restored pulse at the band gap edges can be understood in the frame of the proposed model
assuming the storage of the spin-wave signal in the internal magnonic crystal mode. The efficiency of the excitation of
this mode by the propagating spin wave is of crucial importance. It is clear that this efficiency is determined by the
position of the input signal with respect to the MC's band gaps. When the signal frequency lies between the band gaps
no coupling between the standing and the propagating modes exists. In the opposite situation when the signal frequency
lies inside of the band gap, the propagating wave will be fully reflected and no storage will occur again (the
evanescent wave whose intensity decreases exponential inside the MC area will be formed here). In this case the
spin-wave energy leaves the grooved area in a time interval smaller than the ratio of the double MC's length to the
spin-wave group velocity (under conditions of our experiment $\simeq 230$~ns). Therefore, the only possibility to
effectively excite the standing MC mode is to tune the propagating wave frequency very close to the edge of the band
gap, namely to one of the spin-wave group velocity minima.

Furthermore, the slower the outflow of the spin-wave energy from the area of the parametric amplification (given in
general by the spin-wave group velocity) the higher the amplification rate \cite{Neumann-DR}. That is why the
amplification of the MC standing wave is maximal for the smallest group velocity of the propagating wave which is
responsible for the outflow of the stored energy outside of the magnonic crystal area. This effect contributes
additionally to the strong localization of the restored signals at the band gap edges.

In conclusion, we have demonstrated that spin waves can be stored in magnonic crystals. However, the deceleration of
the propagating spin wave of up to 15 percent is not high enough for the information storage mechanism used in
photonics. In contrary long-time coherent storage has been experimentally realized using the internal standing mode of
the magnonic crystal. By means of further phase-sensitive parametric amplification the stored signal was recovered. The
maximal recovery time was more than $9~\mu$s while the propagation time of the spin waves through the sample was only
0.3~$\mu$s. The results presented here provide deeper understanding of the storage-recovery mechanisms in periodic
lattices in general. Besides, they point to the potential possibility of utilization of magnonic crystals for buffering
or storage of microwave information.

We acknowledge financial support by Deutsche Forschungsgemeinschaft (SE 1771/1-2) and the Australian Research Council.


\begin{thebibliography}{23}
\bibitem{detail1} The group velocity $v_\mathrm{gr}$  has been measured using a network analyzer and the frequency values has been recalculated to the magnetic field. The $v_\mathrm{gr}$ oscillation are due to the beating of the spin-wave signal and the parasitic electromagnetic leakage between the antennas. The calculated $v_\mathrm{gr}$ has been obtained from the phase of the MC transmission matrix. Note that the wave inside of the band gaps is imaginary and the notion of group velocity is not fully applicable to these areas.
\end{thebibliography}
\end{document}